\documentstyle[12pt,epsf]{article}
\input{psfig}
\begin{document}

\def\g{\gamma}
\def\xg{$ x_\g$}
\def\xgo{x_\g^{OBS}}
\def\xgom{x_\g^{cal}}
\def\xglo{x_\g^{LO}}
\def\xp{x_p}
\def\xpo{x_p^{OBS}}
\def\xplo{x_p^{LO}}
\def\pT{p_T}
\def\pZ{p_{\it z}}
\def\Z{{\it z}}
\def\ETC{E_T^{cone}}
\def\ra{\rightarrow}
\def\LE{L_e^{CAL}}
\def\LG{L_{\gamma}^{CAL}}
\def\ETAJ{\eta^{jet}}
\def\ETJ{E_T^{jet}}
\def\ETAP{\eta^{parton}}
\def\ETP{E_T^{parton}}
\def\ETJM{E_T^{cal}}
\def\DETA{|\Delta\eta|}
\def\ETAM{\eta^{cal}}
\def\PHIM{\phi^{cal}}
\def\EEP{E_{e^\prime}}
\def\TEP{\theta_{e^\prime}}
\def\SGP{s_{\gamma p}}
\def\WGP{W_{\gamma p}}
\newcommand{\gsim}{\buildrel{>}\over{\sim}}
\thispagestyle{empty}
\begin{titlepage}
\thispagestyle{empty}

\title{Photon Structure\\ as seen at HERA~\footnote{Presented at the 3rd
Workshop on TRISTAN Physics at High Luminosities, November 16-18, 1994, KEK,
Tsukuba, Japan}}

\author{J. M. Butterworth \\
Dept. of Physics,
Penn State University,\\
University Park, PA 16802, USA.\\
  \\
Representing the ZEUS and H1 collaborations
}

\date{ }

\maketitle

\vspace{5 cm}

\begin{abstract}
At HERA, the electron-proton collider at DESY, Hamburg, the large
flux of almost on-shell photons accompanying the lepton beam is being
used to shed new light on the structure of the photon. Recent results
are reviewed and discussed, with emphasis on those aspects of the photon's
nature which should be understandable using perturbative QCD.
\end{abstract}
\thispagestyle{empty}

\vspace{-19 cm}
{\noindent
\begin{flushright} DESY-95-043,  hep-ph/9503011
\end{flushright} }

\end{titlepage}

\newpage

\section{Introduction}

At DESY, Hamburg, 820~GeV protons collide with 27~GeV electrons
(or positrons) in HERA.
Two general purpose detectors, H1~\cite{h1} and ZEUS~\cite{zeus},
are positioned at opposing interaction points on the accelerator ring.
The high flux of almost on-shell photons which
accompanies the
lepton beam provides a unique opportunity to study the nature of the
photon and its interactions.

Collisions take place between protons and almost on-shell photons (with
a negative mass
squared in the range $10^{-10} < Q^2 < 4$~GeV$^2$), at $\gamma p$
centre-of-mass energies ($\WGP$) up to
300~GeV, an order of magnitude greater than previously available.
HERA does not, of course, measure the structure function $F_2^\gamma$, as
this is defined in terms of the scattering of electrons off photons.
However, studies of jet production in $\gamma p$
interactions provide information about the photon structure
down to momentum fractions as low as $5 \times 10^{-2}$, and because the
photon is being probed
by partons from the proton, there is a high sensitivity to the distribution
of gluons in the
photon.

\begin{figure}[h]
\setlength{\unitlength}{1mm}
\epsfysize=400pt
\epsfbox[50 200 450 600]{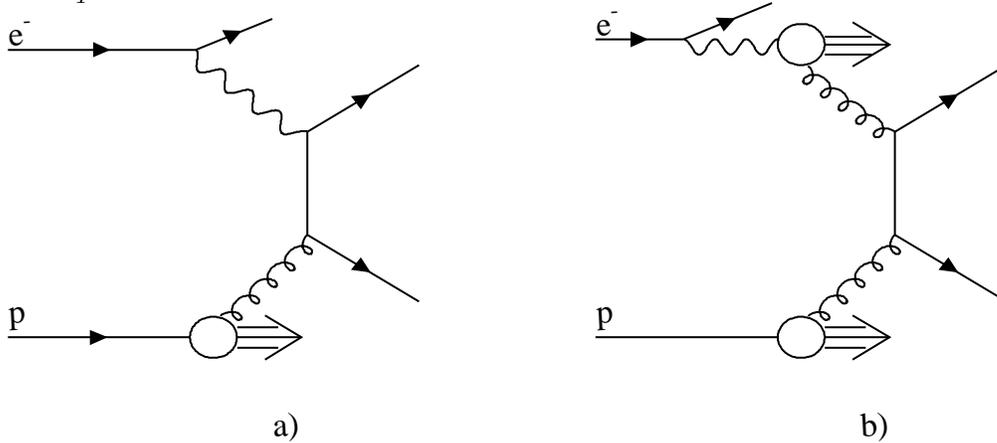}
\vspace{-8cm}
\caption{\label{f:diag}
        { Examples of leading order diagrams for a) direct and b) resolved
photoproduction.}}
\end{figure}

The presence of a `hard' energy scale (usually from the high transverse
energy ($\ETJ$)
of the jets) means that perturbative QCD calculations of event properties
can be confronted
with experiment.
Examples of leading order (LO) QCD diagrams for photoproduction at HERA are
shown in figure~\ref{f:diag}.
At this order, two processes are responsible for jet production. The photon
may interact directly with
a parton in the proton (figure~\ref{f:diag}a), or it may first fluctuate into
an hadronic state (figure~\ref{f:diag}b). In the first case, known as the
direct contribution,
the fraction of the photon momentum ($x_\gamma$) participating in the
hard process is one. The final state consists of two jets,
the proton remnant and the scattered electron.
In the second case, known as the resolved contribution, the photon acts as a
source of partons, which then scatter off partons in the proton, and
$x_\gamma$ is less than one. The final state in this case includes
a photon remnant, continuing in the original
photon direction, in addition to two jets, the proton remnant and the
scattered electron. Direct photon events do not involve the photon structure,
but can probe the parton distributions in the proton down to
$\xp \approx 10^{-3}$,
where $\xp$ is the fraction of the
proton's momentum entering into the hard process. This process is
 directly sensitive to
the gluon distribution in the proton, and complements indirect
extractions~\cite{f2gluon}
using the measurement of $F_2$ in deep inelastic scattering (DIS).

Seen from a perspective of understanding and testing QCD, photoproduction
at HERA provides information about quark and gluon interactions in the
$\gamma p$ system over a broad range of kinematic variables, information
that is complementary to that obtained from DIS, either in $ep$ events at
HERA or in $e\gamma$ events at TRISTAN and LEP.

Both the ZEUS and H1 detectors have high resolution calorimeters and
tracking chambers.
Jets have been measured in the
pseudorapidity\footnote{The $z$ axis is defined to lie along the proton
direction, and $\eta = - $ln$(\tan\frac{\theta}{2}$) where
$\theta$ is the angle between the jet and the $z$ axis.}
region $-1 > \eta > 2.5$ and
with $\ETJ$ in the
range $6 < \ETJ < 37$~GeV.

\section{Untangling the Subprocesses}

\begin{figure}[h]
\setlength{\unitlength}{1mm}
\epsfysize=400pt
\epsfbox[50 50 480 450]{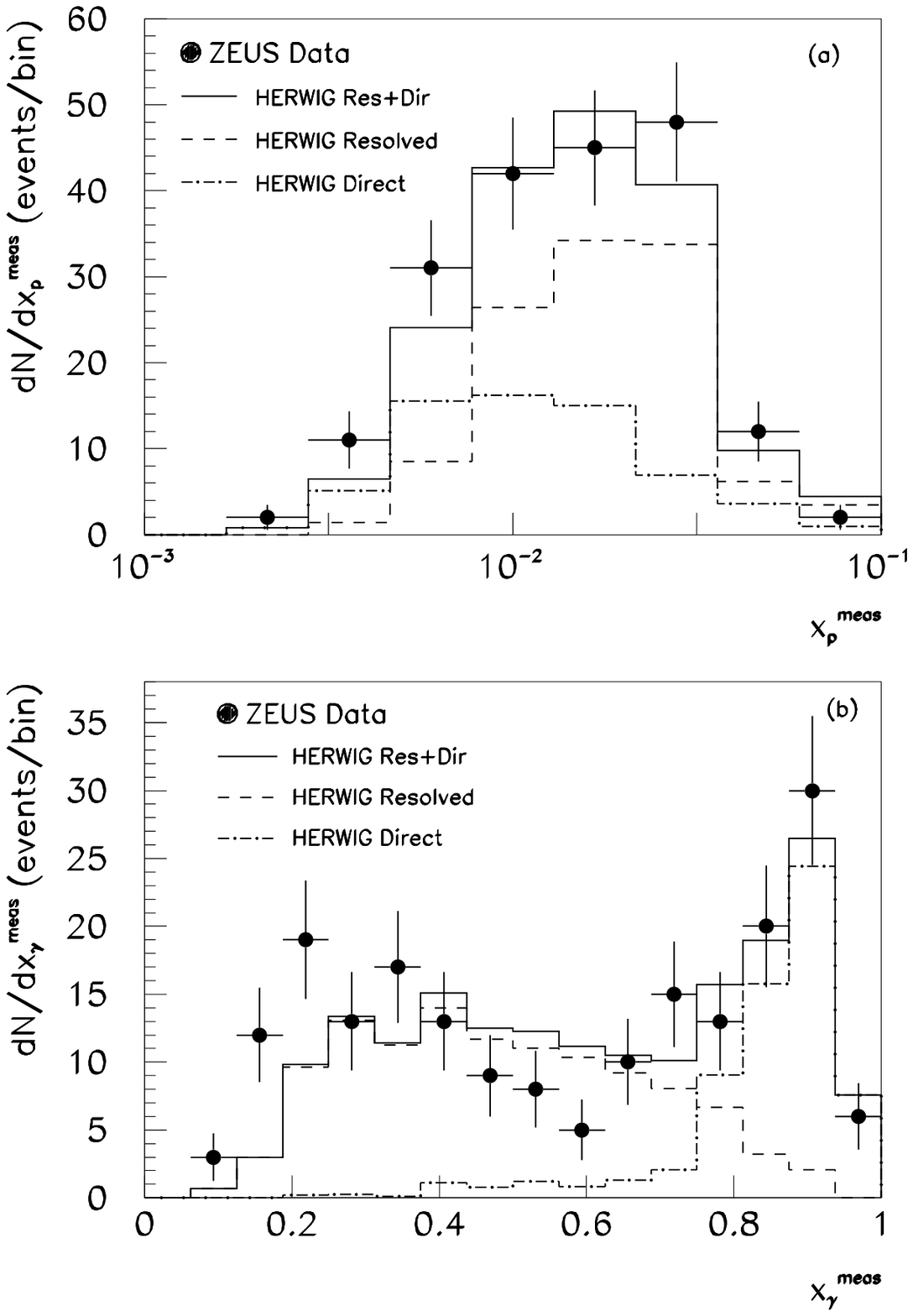}
\vspace{1.0cm}
\caption{\label{f:xg1}
{Kinematic distributions for events with two or more jets.
(a) $x_{p}^{meas}$ ({\it i.e.} uncorrected $\xpo$) distribution for the
final sample.
(b) $x_{\gamma}^{meas}$ ({\it i.e.} uncorrected $\xgo$)
distribution for the final sample.}}
\end{figure}

Because of the large cross sections involved, photoproduction physics provided
interesting results very soon after HERA began colliding the beams. At
that time, evidence for `hard' interactions in the resolved photon process
had just been observed
for the first time, by the AMY collaboration~\cite{amy}. Both the ZEUS and H1
collaborations rapidly provided confirmation
of hard scattering from partons inside the resolved photon, this time
in $\gamma p$ events~\cite{hscat}. Jet production was seen, and energy flow
in the rear (electron) direction provided clear evidence for the presence of
a photon remnant.

In events with two (or more) jets, the parton kinematics of photoproduction
events can be reconstructed. In particular, the momentum fractions of
the photon and proton can be measured. For two-to-two parton scattering in
LO QCD, energy and momentum conservation give the fraction of the photon
energy involved in the hard scatter as

\begin{equation}
\xglo = \frac{ \sum_{partons}E_T^{parton}e^{-\eta^{parton}}}{2yE_e},
\label{xgloeq}
\end{equation}

\begin{equation}
\xplo = \frac{ \sum_{partons}E_T^{parton}e^{\eta^{parton}}}{2E_p},
\label{xploeq}
\end{equation}

\noindent where $yE_e$ is the initial photon energy, $E_p$ is the proton
energy,
and the sum is over the two final state partons. For direct photon events,
$x_\g^{LO} = 1$.
For those events in which two jets are observed
in the detector, it is possible to estimate the fraction of the photon
momentum which participates in the hard process by replacing the sum over
partons
in equation~(\ref{xgloeq}) with a sum over jets.
It is then possible to make an explicit definition of the
observable~\xg,

\begin{equation}
\xgo = \frac{ \sum_{jets}E_T^{jet}e^{-\eta^{jet}}}{2yE_e},
\label{xgoeq}
\end{equation}

\noindent where now the sum runs over the two jets of highest $\ETJ$.
An analogous
definition can be made for the proton momentum fraction.
In the $\xgo$~ distribution thus obtained, the LO direct and resolved
processes
populate different regions, with the direct processes concentrated at
high values
of $\xgo$. The peak arising from the direct contribution will not
necessarily lie exactly
at $\xgo = 1$, due to
higher order effects and/or hadronisation, but will still correspond to the
kinematic region where most or all of the energy of the photon participates in
the hard subprocess. This treatment avoids possible confusion arising due to
the fact the simple separation between resolved and direct processes is
only possible at LO in QCD and $\xglo$~ becomes ambiguous even at
next-to-leading order (NLO).

The $\xgo$ and $\xpo$ distributions (not corrected for detector effects, and
therfore called $x^{meas}$) measured by the ZEUS collaboration using 1992
data~\cite{direct} are shown in
figure~\ref{f:xg1}. The $\xgo$ distribution (figure~\ref{f:xg1}b) rises at
both low and high values. The Monte Carlo simulations of the resolved and
direct processes (which also include detector effects) shown in the same
figure, have very different
characteristics. The resolved processes show a rise towards low $\xgo$,
as observed in the data, but cannot account for the rise at high $\xgo$.
The direct processes predict a sharp rise towards high $\xgo$ as observed in
the data and only a small number of events for $\xgo < 0.7$.
The conclusion is that the peak at the high end of the $\xgo$ distribution
results from direct processes.

\section{Jet Cross Sections}

The results presented in the previous section
show good qualitative agreement between the data and QCD
expectations, in the
sense that the expected production mechanisms and event topologies are
observed. However,
signs of discrepancies are already seen. In particular, when the transverse
energy flow around the jets
is studied, more energy was seen in the data in the forward (proton)
\clearpage
region
than was expected
from Monte Carlo simulations~\cite{h1ped}.
In addition, in figure~\ref{f:xg1}
there is an excess of low $\xgo$ events in the data over the simulation.
In order to obtain quantitative information from photoproduction processes,
clearly defined
cross sections must be measured and compared with theoretical calculations.
I concentrate here on jet cross sections, although much information can be
also be obtained from measurements of charged particle
spectra~\cite{cps}.

\subsection{Inclusive Jet Cross Sections}

Inclusive jet cross sections provide a good test of perturbative QCD,
and have been measured by both the H1~\cite{h1ped} and ZEUS~\cite{zinc}
collaborations.
At both experiments, jet finding is performed
using a  cone algorithm~\cite{snow}. The cone radius of
the jet finding algorithm is typically defined such that
$R = \sqrt{\Delta\eta_{cell}^2 + \Delta\phi_{cell}^2} = 1$, although
other values could also be used in the future. The algorithm searches in
pseudorapidity-azimuth
($\eta_{cell}$-$\phi_{cell}$) space for the cone containing the highest
summed $E_T$.

H1 use the electron calorimeter of their luminosity monitor~\cite{h1ped}
to measure the energy of the low angle scattered electron in the range
$0.25 < y < 0.7$, where $y = E_\gamma/E_e$, the fraction of the
initial electron energy ($E_e$) carried by the almost real photon with
energy $E_\gamma$. The corresponding range of photon virtualities
is approximately $10^{-8}$~GeV$^2 < Q^2 < 0.01$.
For jet studies at ZEUS, photoproduction events are usually defined by
demanding that the
electron is scattered at small angles and does not emerge from the beam pipe
with the central ZEUS detector.
This requirement corresponds approximately to a cut of
$Q^2 < 4$~GeV$^2$. The range of $y$ values used by ZEUS is $0.2 < y < 0.85$.

\begin{figure}[h]
\setlength{\unitlength}{1mm}
\epsfysize=400pt
\epsfbox[50 50 500 450]{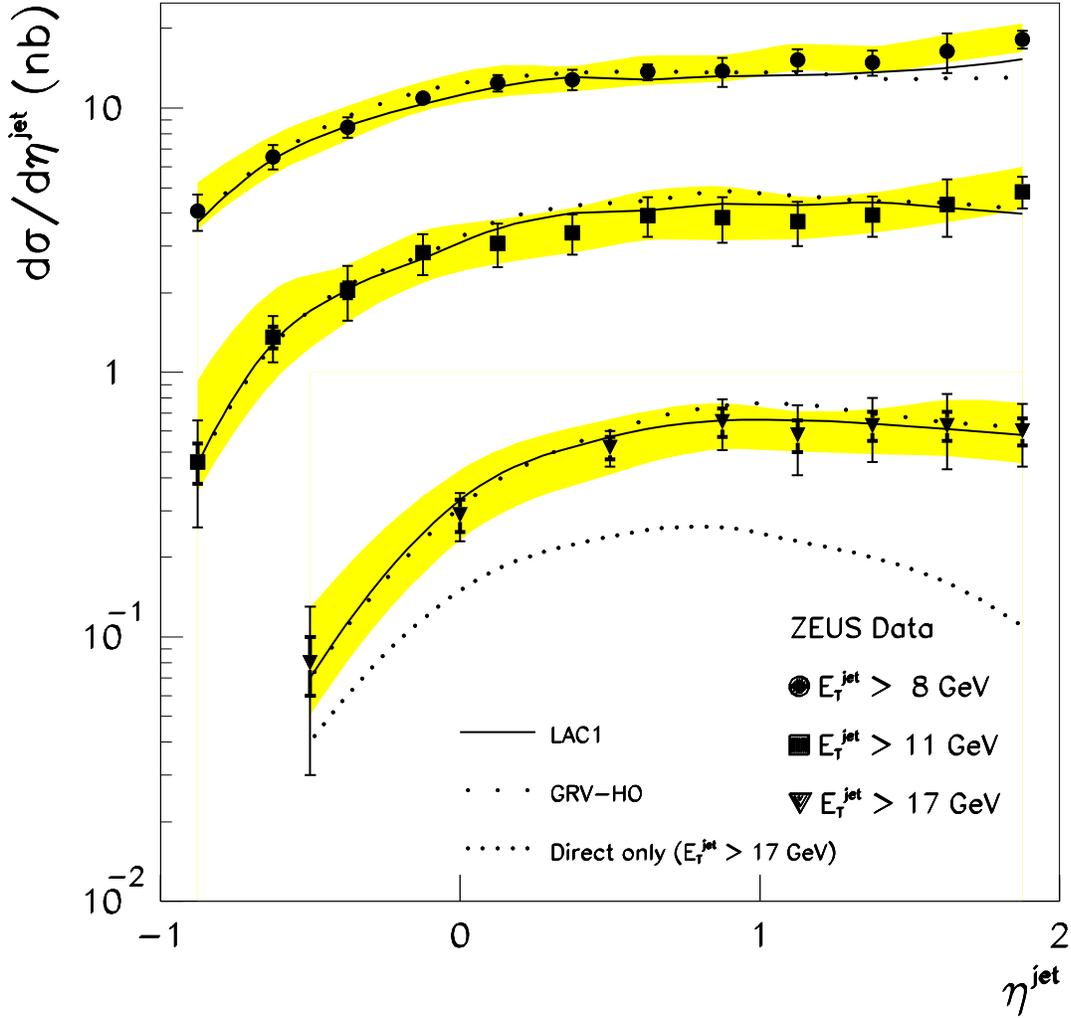}
\vspace{1.0cm}
\caption{\label{f:inceta}
         { Differential $ep$ cross section $d\sigma/d\eta$ for inclusive jet
production, integrated over $\ETJ$ from three different thresholds
($\ETJ > 8, 11$ and $17$~GeV) in the kinematic region defined by
$Q^2 \le 4$~GeV$^2$ and $0.2 > y > 0.85$. PYTHIA calculations are shown for
comparison.}}
\end{figure}

\begin{figure}[h]
\setlength{\unitlength}{1mm}
\epsfysize=400pt
\epsfbox[50 50 500 450]{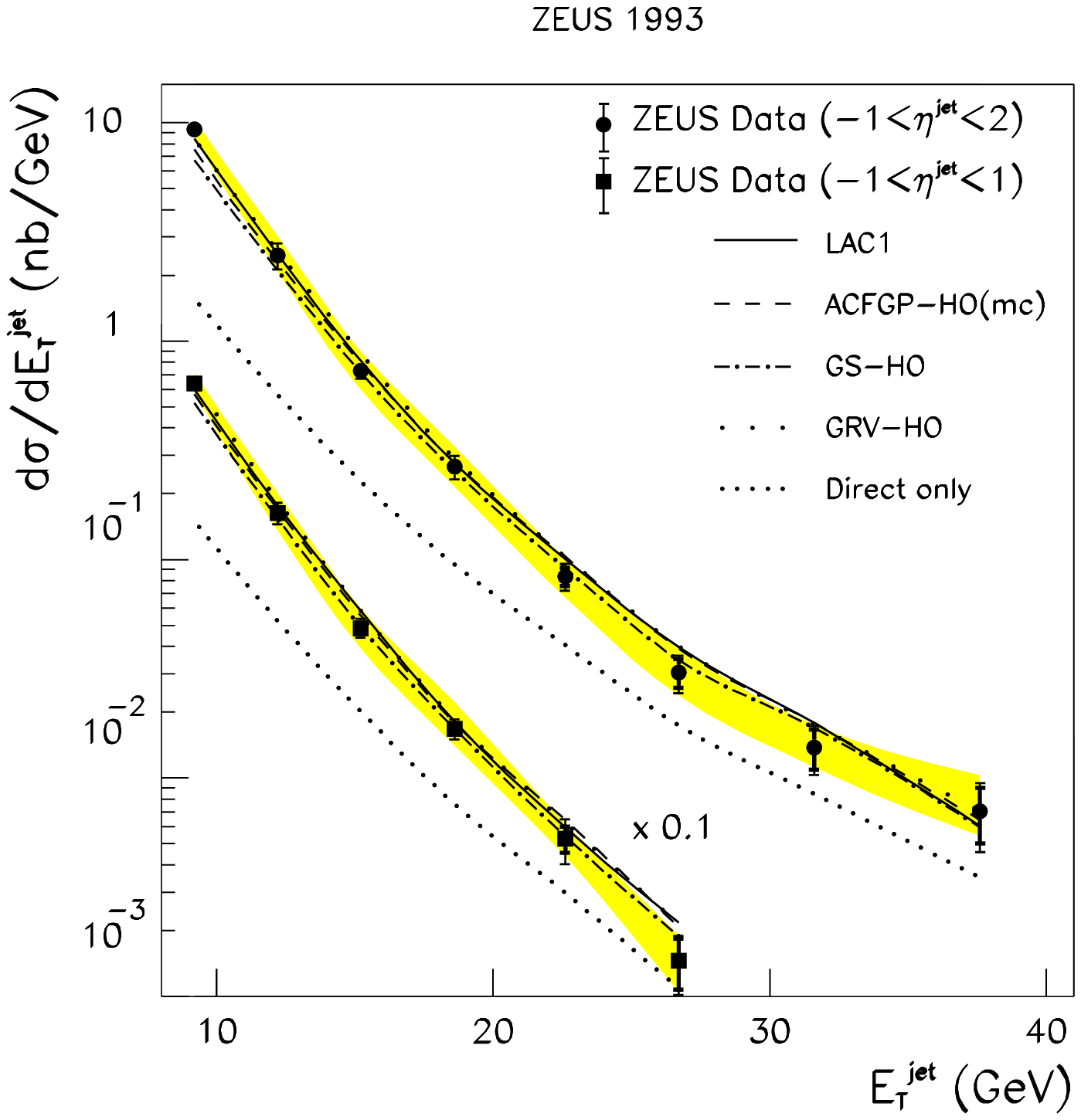}
\vspace{1.0cm}
\caption{\label{f:dsdet}
{Differential $ep$ cross section $d\sigma/d\ETJ$ for inclusive jet
production, integrated over
two $\ETAJ$ ranges ($-1 < \ETAJ < 2$ and $-1 < \ETAJ < 1$) in the kinematic
region defined by $Q^2 \le 4$~GeV$^2$ and $0.2 > y > 0.85$.
PYTHIA calculations are shown for comparison.
For the latter $\ETAJ$ range, both Monte Carlo curves and data have
been multiplied by 0.1.}}
\end{figure}

The inclusive jet cross cross section $d\sigma/d\eta$
measured by the ZEUS collaboration is shown in figure~\ref{f:inceta}.
Also shown are
the cross sections calculated by the PYTHIA~\cite{PYT} Monte Carlo model
(including the effects of hadronisation and
parton showering). The model with both the photon
parton distribution sets shown describes the data well,
with only a small discrepancy in the foward (high $\eta$) region for
the low $\ETJ$ cross section.
The inclusive jet cross section $d\sigma/d\ETJ$ is shown in
figure~\ref{f:dsdet}, for two different ranges of $\ETAJ$, one of which
includes the forward region and the other of which excludes it.

Given the failure of the Monte Carlo models used here to describe the
energy flow in the forward region shown in ref.~\cite{h1ped}) it is not
surprising that the cross sections do not completely agree in the forward
region at low $\ETJ$. At higher $\ETJ$ the uncertainties arising from
hadronisation can be expected to become less significant, and any test of
QCD {\it via} such
comparisons becomes less ambiguous. NLO calculations of these processes are
already becoming available~\cite{NLO}.

\clearpage

\subsection{Dijet Cross Sections}

Although inclusive jet cross sections have the advantage of
calculational simplicity, there are advantages in measuring
dijet cross sections. In this case, reconstruction of the kinematics
of the hard scatter becomes possible, in particular of $\xgo$ and $\xpo$.
By making kinematic cuts it is possible to untangle some of the many
unknowns in the photoproduction process, in particular the dependence upon
the parton distributions in the proton and photon.

The ZEUS collaboration has measured the differential cross section
$d\sigma/d\bar{\eta}$
for dijet photoproduction~\cite{dijets} \\
($ep~\rightarrow~e\gamma~p~\rightarrow~e~X~+~\mbox{2 jets}$)
in the $y$ interval  $0.2 < y < 0.8$, for events
with at least two jets of $\ETJ > 6$~GeV,
and a difference in the pseudorapidities
of the two jets of $\DETA < 0.5$.
The variable $\bar{\eta} = \frac{1}{2}(\eta_1 + \eta_2)$ is the average
pseudorapidity of the two jets of highest transverse energy.
This cross section is measured separately for resolved and direct photon
events, as defined by a cut on $\xgo$ at 0.75, with the low $\xgo$ events
identified with resolved photoproduction and the high $\xgo$ events identified
as direct photoproduction.
The expression for $\xgo$ (equation~\ref{xgoeq}) can be rewritten in
terms of $\DETA$ and
$\bar\eta$, assuming the jets to have equal transverse energy, to obtain
\begin{equation}
\xgo = \frac{E_T^{jet}e^{-\bar{\eta}}}{yE_e} \cosh \frac{\Delta \eta}{2}.
\label{xgoeq2}
\end{equation}
A similar expression can be written for the proton,
\begin{equation}
\xpo  = \frac{E_T^{jet}e^{\bar{\eta}}}{E_p} \cosh \frac{\Delta \eta}{2}.
\label{xgoeq3}
\end{equation}
Constraining the jets to be at almost equal pseudorapidities, with the
cuts on $\DETA$, means that
the hyperbolic cosine term takes its minimum value of unity.
Thus the minimum available
$x$ values are probed for a given $\ETJ$ and $\bar{\eta}$, and there
is a strong correlation between $\bar{\eta}$ and $\xpo$ in the direct cross
section and between $\bar{\eta}$ and $y\xgo$ in the resolved cross
section~\cite{jeff1}.

\begin{figure}[h]
\setlength{\unitlength}{1mm}
\epsfysize=400pt
\epsfbox[50 350 450 750]{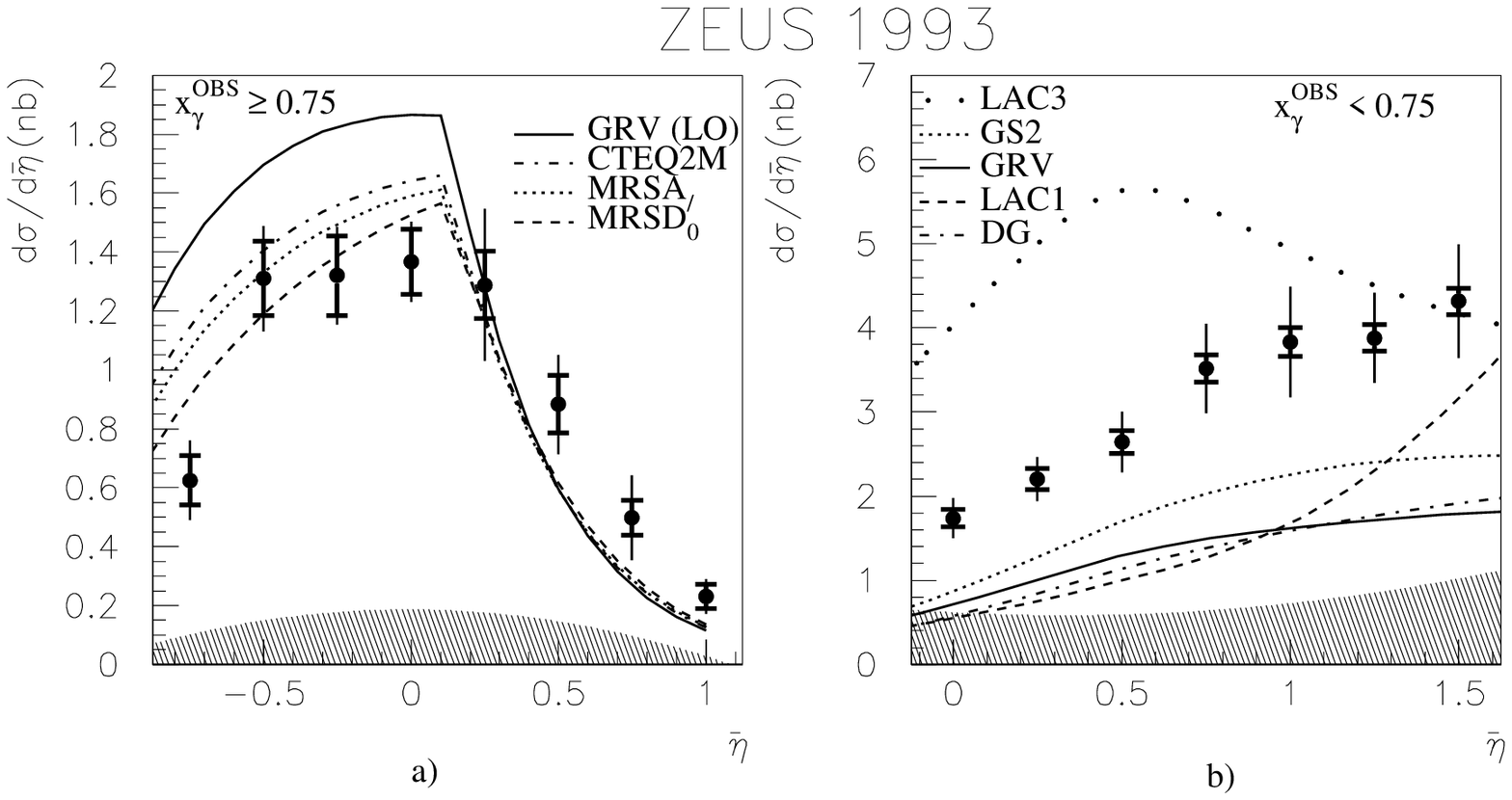}
\caption{\label{f:dij}
           { $d\sigma/d\bar{\eta}$ for $ep \rightarrow eX + \mbox{2 jets},
\DETA < 0.5, \ETJ > 6$~GeV,
$0.2 < y < 0.8, Q^2 < 4$~GeV$^2$,
a) $\xgo \ge 0.75$, b) $\xgo \ge 0.75$,
The solid circles are corrected ZEUS data.
The inner error bars indicate the
statistical errors, the outer error
bars show the systematic uncertainty excluding the
the correlated uncertainty from measurement of
energy in the calorimeter and the integrated luminosity, which are
show as a shaded band.
In a) the data are compared to LO QCD calculations using several parton
distribution sets for the proton and the GS2 set for the photon.
In b) the data are compared to LO QCD calculations using
the LAC3, GS2, GRV, LAC1 and DG parton distribution for the
photon. The proton parton distribution set used is the MRSA set.}}
\end{figure}

The direct cross section is shown in figure~\ref{f:dij}a, compared to
leading order QCD calculations using various different parton distribution
sets for the proton. The parton distribution sets used are the
GRV LO~\cite{GRV}, CTEQ2M~\cite{CTEQ}, MRSA~\cite{MRSA} and
MRSD$_0^\prime$~\cite{MRS}
parton distribution sets for the proton.
The shape of the direct cross section differs from that
of the LO QCD calculations. However, several effects
(apart from the choice of parton distribution set) influence the comparison
between data and theory. QCD calculations are only available at LO, and higher
order corrections may be large. Also, due to the fact that we are probing
low $x$ partons in the proton, the standard approximation that the
incoming partons are collinear and on-shell may be invalid~\cite{jeff2}.
Finally, non-perturbative `hadronisation' effects can be
expected to be significant.
Considering these uncertainties the agreement with the theory is fairly
good.

\clearpage

Figure~\ref{f:dij}b shows the measured cross section
$d\sigma/d\bar{\eta}$ for resolved photoproduction.
The LO cross sections, shown for comparison, are calculated using different
photon parton distribution sets and the MRSA~\cite{MRSA} set
for the proton. The theoretical sensitivity to the parton distributions in
the proton is small (not shown), with the variations between curves calculated
using different parton distribution sets being much less than the estimated
errors on the measured cross section.
The DG~\cite{DG}, GRV~\cite{GRV} and GS2~\cite{GS} parton distribution sets
reproduce the shape
of the cross section well and can be brought into agreement with the data by
applying a multiplicative factor of around 2. Factors of this size may
reasonably be expected to come from NLO calculations~\cite{NLO}.
The LO calculations using the
LAC1 and LAC3~\cite{LAC} parton distribution
sets cannot be brought into agreement with the data by a constant
normalization factor.

\section{Photon Structure}

The Monte Carlo simulations discussed so far attempt
to describe photoproduction by a hard parton-parton subprocess
coupled with parton distributions in the proton and photon and parton
shower and hadronisation models. No allowance is made for the possibility
of further interactions (other than a single parton-parton scattering)
taking place in a $\gamma p$ event. In obtaining the results shown, the
simulations are only used to correct for detector
acceptance and smearing. The effects of the discrepancies between the
Monte Carlo simulations and the data on the measurement of the cross sections
are estimated and included in the systematic errors. Thus the cross sections
can be presented as measurements of observed jet cross sections with
confidence. However, this confidence evaporates when we come to interpret
the cross sections in terms of parton distributions. If there are unknown
other processes contributing to the energy flow in the event, they will
distort the correlation between `parton' kinematic variables in the model
and jet variables in the data.

\begin{figure}[h]
\setlength{\unitlength}{1mm}
\epsfysize=400pt
\epsfbox[0 200 450 650]{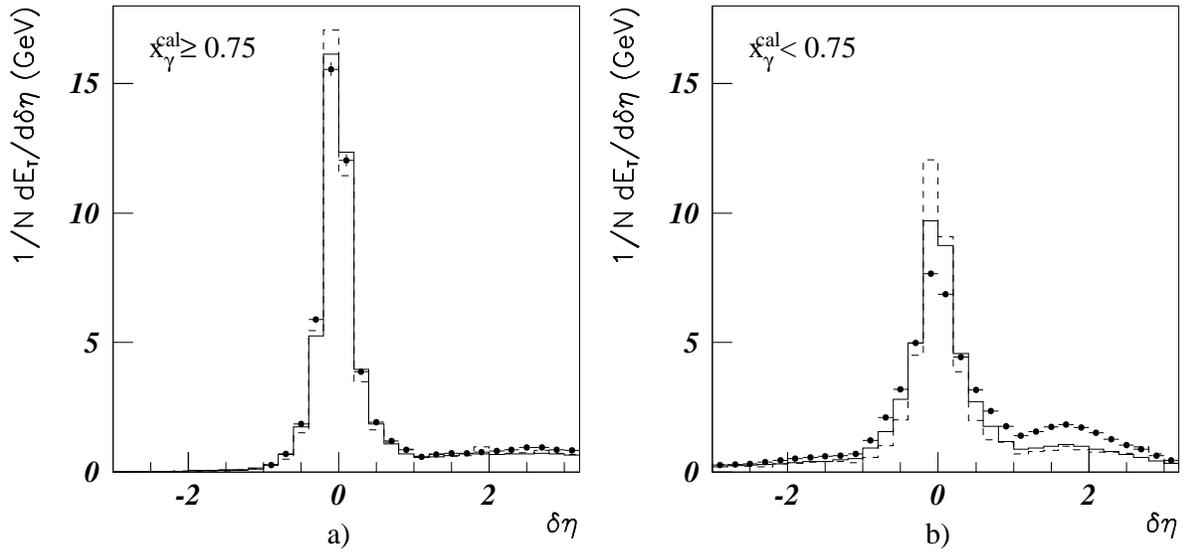}
\caption{ \label{f:profs}
{Figures~a) and b) show the uncorrected transvserse energy
flow $dE_T/d\delta\eta$ around the jet axis,
for cells within one radian in $\phi$ of the jet axis, for a) direct
and b) resolved events. The solid (dashed) line represents the
distribution from PYTHIA (HERWIG).}}
\end{figure}

As an example of this, figure~\ref{f:profs} shows the
uncorrected transverse energy flow per jet
$1/N dE_T/d\delta\eta$ around the jet axis for events entering the two dijet
cross sections of figure~\ref{f:dij}. For the events with
$\xgom \ge  0.75$ (direct events, figure~\ref{f:dij}a and
figure~\ref{f:profs}a), both the HERWIG~\cite{HRW} and PYTHIA
simulations reproduce the data distribution well.
The same distribution is shown in
figure~\ref{f:profs}b for events with $\xgom < 0.75$ (resolved).
In this case both simulations fail to describe the pedestal in the
energy flow in the forward
region, as was also observed (for inclusive jets without an $\xgo$ cut)
in references~\cite{h1ped} and~\cite{zinc}.
Thus when interpreting the cross section in
figure~\ref{f:dij}b in terms of parton distributions,
it is difficult to estimate what the
effects of hadronisation and parton
showering might be on the shapes of the curves, since the jet
pedestal energies
are not well described in the Monte Carlo.
Therefore no parton distribution set (with the exception of LAC3)
can be confidently excluded at this stage.

\clearpage

\subsection{Multiple Interactions}

One candidate for improving the agreement between the Monte Carlo
models and the data is to allow for the possiblity of more than one
hard interaction per $\gamma p$ event~\cite{mi}. This can happen
because the parton densities probed at HERA are sufficiently large that even
from a naive probabilistic point of view, the chance of more than one hard
scatter in a $\gamma p$ interaction becomes significant. Models of this nature
are usually called `eikonal' models and are also used to regularise the strong
growth exhibited in minijet calcuations of the $\gamma p$ cross
section~\cite{eik}. It has been demonstrated by H1~\cite{h1glasgow} that
including some form of multiple interaction model can bring the Monte Carlo
simulation into significantly better agreement with the data.
As multiple interactions
in this type of scenario are generated by the density of partons in the photon,
they are expected to be absent in the direct process, where the photon parton
distributions play no part.
Returning to the jet profiles in the resolved and direct samples isolated by
ZEUS in reference~\cite{dijets} (figure~\ref{f:profs}) the fact
that the forward energy excess is not observed in the
direct (high $\xgo$) events lends support to the idea that multiple hard
interactions are a reality.

\subsection{Parton Distributions}

\begin{figure}[h]
\setlength{\unitlength}{1mm}
\epsfysize=400pt
\epsfbox[20 0 470 200]{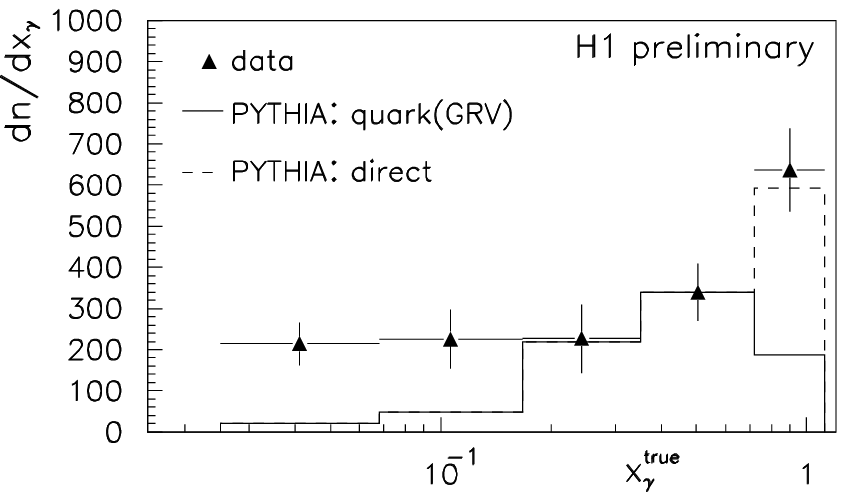}
\vspace{1.0cm}
\caption{\label{f:h1xg}
         { The $\xglo$ distribution unfolded using PYTHIA from events with
two jets $\ETJ > 7$~GeV and $\ETAJ < 2.5$.}}
\end{figure}

It is possible to use the PYTHIA multiple interaction model, together with an
additional `underlying' jet pedestal energy and the usual parton showering and
hadronisation models, to obtain a good description of the
energy flow around the jets~\cite{h1glasgow}. Within this model,
there then exists a particular correlation between
the LO, generated partons and the observable jets.
In the absence (so far) of better theoretical tools, it is interesting
to assume that the Monte Carlo model represents the `truth' as far as this
correlation is concerned, and deduce what parton distribution is needed in
the Monte Carlo simulation in order to successfully describe the observed
jet rates. This procedure was carried out by H1, and the resulting
$\xglo$ distribution is shown in figure~\ref{f:h1xg}.

\begin{figure}[h]
\setlength{\unitlength}{1mm}
\epsfysize=400pt
\epsfbox[20 0 470 200]{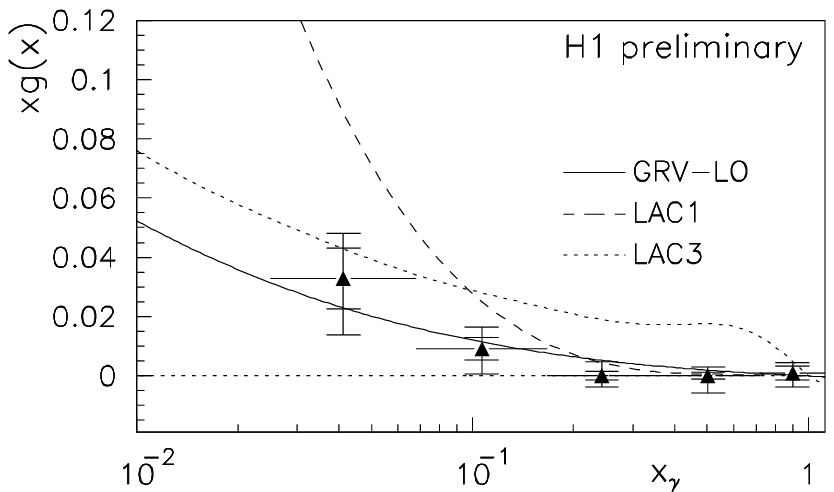}
\vspace{1.0cm}
\caption{\label{f:h1gl}
         { LO estimate of the gluon density in the photon unfolded
using PYTHIA from events with two jets $\ETJ > 7$~GeV and $\ETAJ < 2.5$.}}
\end{figure}

Once this has been done, the fact that measurements of $F_2^\gamma$
constrain the quark distributions in the
photon in this kinematic region can be
used to deduce what part of the parton density is attributable to gluons in
this model. The result is shown in figure~\ref{f:h1gl}.

\subsection{Remnant Structure}

Most analyses of photoproduction events at HERA have implemented
a cone algorithm to find jets. The cone algorithm uses a cone of
fixed radius in pseudorapidity and azimuthal angle space and is well
suited for high transverse energy jets at fairly central pseudorapidities.
However, the photon remnant will mainly deposit energy at low angles
relative to the electron direction, over a large negative
pseudorapidity range, and a clustering algorithm
is more appropriate. ZEUS have used the $k_T$ jet clustering algorithm
in the laboratory frame~\cite{kt,ktep} to isolate and study this photon
remnant~\cite{rem}.

This algorithm finds jets by calculating a quantity $k_T$ for each
calorimeter cell `hit' with respect to all other hit cells.
In the small angle approximation, $k_T$ reduces to the
transverse momentum squared of the lower energy cell with respect to the
higher
energy cell. The value of $k_T$ for each pair of cells, or clusters, is
calculated from the formula:
\begin{equation}
k_T = 2\mbox{min}(E^2_n,E^2_m)(1-\cos\theta_{nm}),
\end{equation}
where $E_n$ and $E_m$ are the energies for cells $n$ and $m$, and
$\theta_{nm}$
is the angle between the two cells. When all of the $k_T$ values
of an event have been calculated, the two cells with
the lowest $k_T$ value are merged.  This process is iterated, merging cells
or clusters of cells, until the desired number
of clusters remain. The proton remnant is included in
the clustering procedure as a
pseudo-particle which moves in the proton direction.
For high transverse energy jets the K$_t$ cluster finding algorithm
gives results similar to those obtained with the usual cone algorithm.

\begin{figure}[h]
\setlength{\unitlength}{1mm}
\epsfysize=300pt
\epsfbox[0 100 450 500]{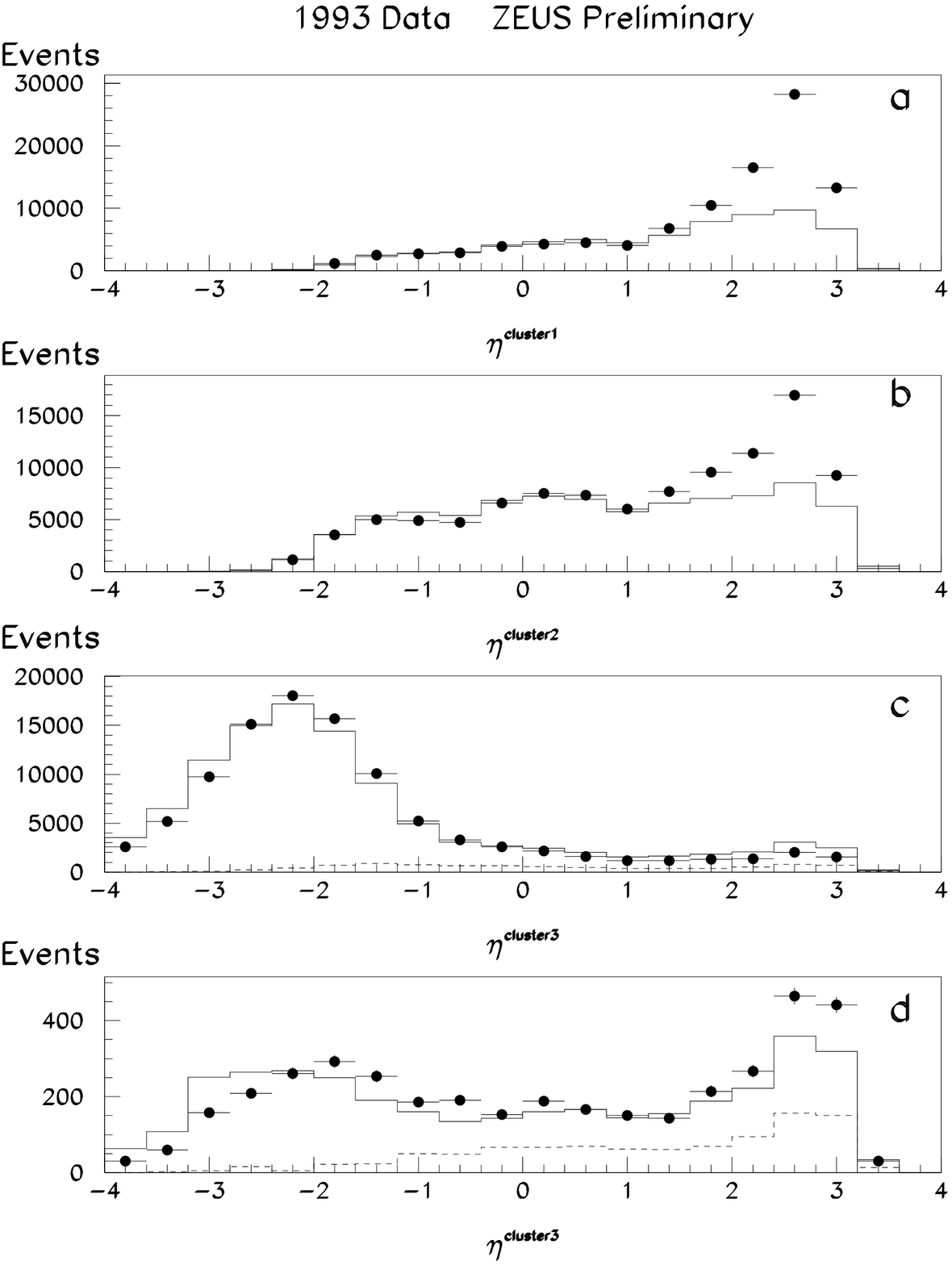}
\vspace{2cm}
\caption{\label{f:rem1} Pseudorapidity
distributions of the first (a), second (b)
and third (c) clusters found by the K$_T$ algorithm: ZEUS data (full
circles), MC direct+resolved (full histogram) and MC direct
(dashed histogram). (d) as (c) but imposing a cut at $P_T^{cluster_{1,2}}
\ge 5.0$ GeV and $\eta^{cluster_{1,2}} \le 1.6$. Data and MC events are
normalized in the region $\eta \le 1.6$}
\end{figure}

The $k_T$ algorithm was forced to find three energy clusters in each
event, in addition to the cluster related to the proton remnant.
The pseudorapidity $\eta^{cluster}$ distributions of the three
clusters obtained with the $k_T$ algorithm are shown in figure~\ref{f:rem1}a-c,
in order of decreasing transverse energy.
The data ( full circles) and MC ( resolved and direct, histogram) are shown
normalized in the region $\eta^{cluster} \le 1.6$.
While the first and second clusters are predominantly found in the
$\eta^{cluster} \ge 0$ region, the third cluster
has an accumulation in $\eta^{cluster} \le 0$,
{\it i.e.} the outgoing electron direction.
The agreement between the data and the MC is reasonable for all three
clusters except for the very forward region $\eta^{cluster} \ge 1.6$
where the discrepancies already noted in the forward energy flow can be
expected to be significant. In particular
the excess observed in figure~\ref{f:rem1}c in the negative pseudorapidity
region is well described by the Monte Carlo simulation including resolved
and direct processes (full histogram).
However the (LO) direct process does not contribute to this
excess as shown by the dashed line in the figure; its distribution
is rather flat as expected from the absence
of photon remnants for this class of events. As a consequence,
the third cluster in the negative pseudorapidity region
can be identified with the photon remnant.

After requiring the first two clusters to be jets with $\ETJ > 5$ GeV
and pseudorapidity $\ETAJ < 1.6$, as in reference~\cite{direct},
the $\eta$ distribution of the third cluster is shown in figure 1d.
Compared with the MC expectations, the data show an excess
toward higher $\eta$ values corresponding to
higher angles with respect to the electron direction.
A cut of $\eta^{cluster_3} < -1$ was then applied to the third cluster.
After these cuts, $\xgo$ was found to peak at low
values as expected for resolved processes.

\begin{figure}[h]
\setlength{\unitlength}{1mm}
\epsfysize=300pt
\epsfbox[0 100 450 500]{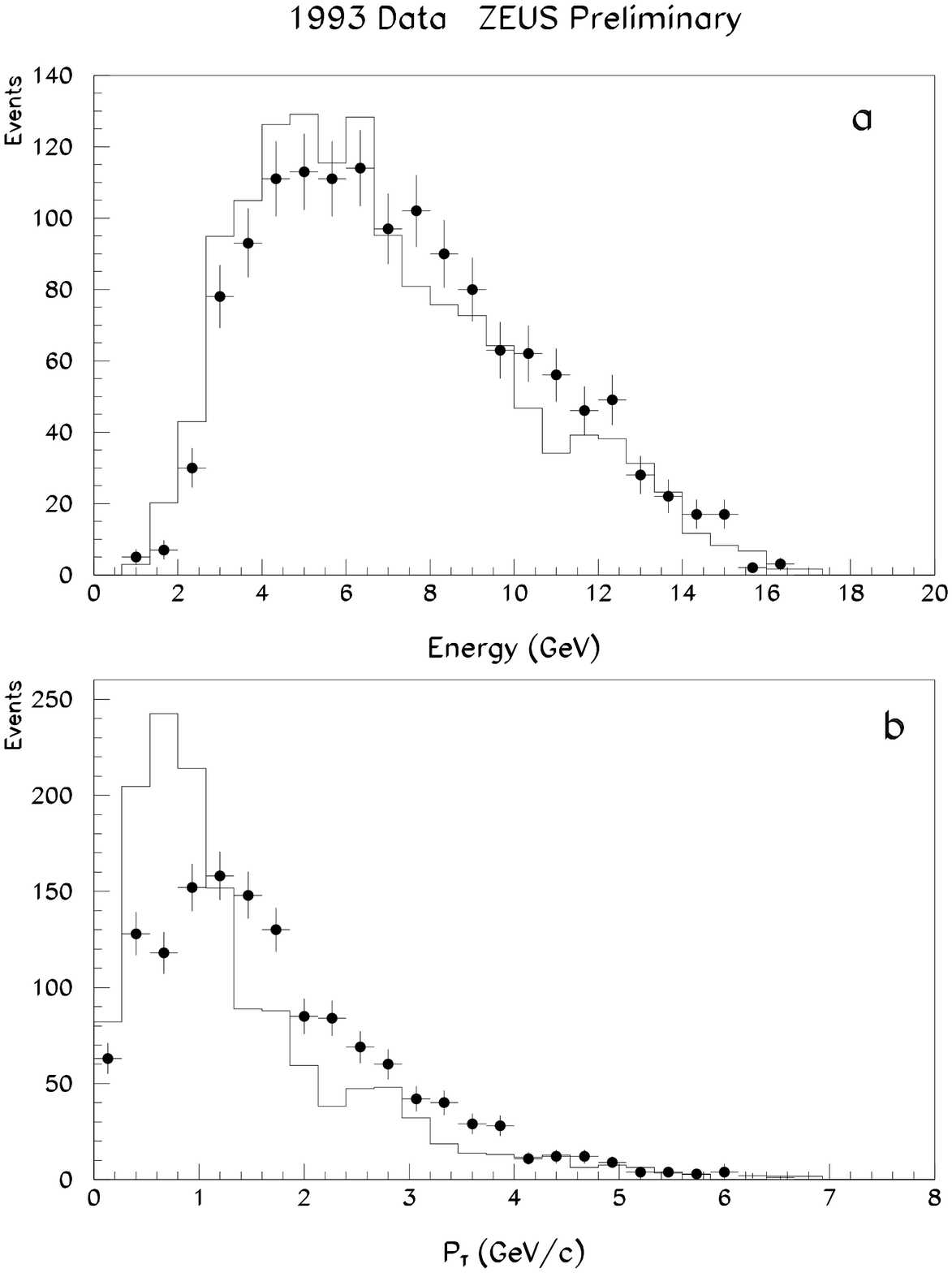}
\vspace{2cm}
\caption{\label{f:rem2} Energy (a) and $P_T^{cluster}$ (b)
distributions for third cluster:
ZEUS data (full circles) and MC direct+resolved (full histogram).
Data and MC are normalized to the number of events in data.}
\end{figure}

The energy distribution for the third cluster is shown in figure~\ref{f:rem2}a
The solid histogram shows the Monte Carlo expectation which
agrees reasonably with the data.
The transverse momentum (with respect to the beam) distribution of the
third cluster, shown in figure~\ref{f:rem2}b, peaks at 1 GeV
with a tail extending to 6 GeV.
The Monte Carlo expectation, shown in histogram,
shows a lower average value. This is in qualitative agreement with
the expectation is that the `anomalous' component of resolved
photoproduction, where the photon splits into a perturbatively calculable
$q\bar{q}$ pair, should have a higher transverse energy remnant than the
so-called Vector-meson (VMD) like component, where the splitting is
non-perturbative. The Monte Carlo models so far used assume the
VMD-like topology for all events.

\section{Summary}
HERA data provides unique opportunities to study the structure and
interactions of the photon. Jet production by both the whole photon and
its parton constituents have been unambiguously observed, and jet cross
sections are being used to test QCD and constrain the parton (particularly the
gluon) distributions in the photon and the proton. Studies of the
structure of the photon remnant give indications of a higher transverse
remnant energy than is expected in simulations which describe the photon as
a vector meson.

\section*{Acknowledgements}
I am grateful to DESY and to KEK for assistance with travelling expenses,
and to colleagues in the ZEUS and H1 collaborations for providing the
material presented. This work was partially funded by the US National
Science Foundation.


\end{document}